# Oblivious Transfer based on Key Exchange

Abhishek Parakh

*Abstract:* Key-exchange protocols have been overlooked as a possible means for implementing oblivious transfer (OT). In this paper we present a protocol for mutual exchange of secrets, 1-out-of-2 OT and coin-flipping similar to Diffie-Hellman protocol using the idea of obliviously exchanging encryption keys. Since, Diffie-Hellman scheme is widely used, our protocol may provide a useful alternative to the conventional methods for implementation of oblivious transfer and a useful primitive in building larger cryptographic schemes.

## 1 Introduction

Oblivious transfer (OT), discussed by Stephen Wiesner as conjugate coding [1] became popular when Rabin described a scheme for mutual exchange of secrets [2]. This combined with 1-out-of-2 oblivious transfer led to the development of numerous cryptographic tools.

An oblivious transfer protocol is a scheme in which Alice transfers to Bob a secret without knowing if Bob received it, while Bob may or may not receive the secret, each happening with a certain probability, usually one-half. Such a scheme using Elliptic Curve Cryptography has been discussed in [3].

In this paper we construct a protocol for oblivious transfer using key exchange similar to Diffie-Hellman (DH) protocol [4], which is a popular method for establishing a shared key between two parties over an insecure channel. We modify the Diffie-Hellman protocol such that the two communicating parties will succeed or fail in establishing a shared key each with a probability of one-half. However, the party sending the secret will not know if the receiver has the same key as he/she does.

There have been implementations [5,6] of 1-out-of-n OT based on the Decision Diffie-Hellman (DDH) problem [7]. However, our protocol differs from previous ones in the sense that - firstly, we describe a scheme for mutual exchange of secrets based on DH. Secondly, in the previous implementations the 1-out-of-n OT use the DDH for the transfer itself, i.e. applies the Diffie-Hellman exponentiation for the encryption of secrets directly. Here we administer the idea of the oblivious key exchange. Once the keys are exchanged (obliviously), the parties may use any mutually agreed encryption method for the actual transfer / exchange of secrets.

The security of our protocol arises from the fact that the problem of determining an exponent $e$ given $x, y$ and a prime $p$, such that $x^e \bmod p = y$ is equivalent to solving a Discrete Log Problem (DLP) efficiently.

The participants choose numbers $p$ and $x$, such that $p$ is a large prime on the order of at least 300 decimal digits (1024 bits), $p-1$ has a large prime factor and $x$ is a generator of order $p-1$ in the multiplicative group $Z_p$ (a generator is a primitive root of $p$). This ensures the security of the protocols not only against eavesdroppers but also against the opposing party, which is to be considered as an adversary as well. Since we will be working only in $Z_p$, we often do not state it explicitly.

The proof that the security of Rabin's cryptosystem is equivalent to a factorization problem led to the development of the zero-knowledge proof [8]. In such a proof a prover tries to convince a verifier that he possesses certain information but he does not disclose the information but only the proof that he possesses the information. With every iteration of the algorithm, the probability of an imposter cheating a verifier decreases exponentially. We will discuss a scheme for zero-knowledge proof based on the discrete log problem.

## 2 Mutual exchange of secrets

Suppose Alice and Bob possess secrets $S_A$ and $S_B$ respectively, which they wish to exchange, however, they do not trust each other. We would like to complete the exchange without a trusted third party and without a procedure for simultaneous exchange of secrets; the latter being practically impossible to implement when the parties are geographically far apart. Both parties are assumed to have an appropriate mechanism to digitally sign every message they send.

Let the secrets $S_A$ and $S_B$ be passwords to files that Bob and Alice want to access such that if a wrong password is used then the files will self-destruct. This prevents the parties from trying random passwords. The protocol is based on the oblivious exchange of encryption keys.

***The Protocol***: We exploit the fact that there exist $g_1, g_2 \in Z_p$, $g_1 \neq g_2$ such that they map to a single cipher $c$, where $c = g_1^2 \bmod p = g_2^2 \bmod p$. Let $K_A$ denote the key that Alice uses to encrypt her secret, while Bob uses $K_B$ to encrypt his secret. With these assumptions, the protocol proceeds as follows:

1. Alice and Bob agree upon a prime $p$, a number $x \in Z_p$ as the generator and $c$ such that $c = g_1^2 \bmod p = g_2^2 \bmod p$ (Alice and Bob both know $g_1$ and $g_2$).

2. Alice privately chooses $g_A = g_1$ or $g_A = g_2$ and two random numbers $N_{A_1}$ and $N_{A_2}$.

3. Bob secretly decides on $g_B$, such that $g_B = g_1$ or $g_B = g_2$ and a random number $N_B$.

4. Alice sends to Bob: $x^{g_A+N_{A_1}} \mod p$ and $x^{N_{A_2}} \mod p$.

5. Bob sends to Alice: $\left(\dfrac{x^{g_A+N_{A_1}}}{x^{g_B}}\right)^{N_B} \mod p$ and computes $K'_A = \left(x^{N_{A_2}}\right)^{N_B} \mod p$ for himself.

6. Alice computes: $K_A = \left[\left(\dfrac{x^{g_A+N_{A_1}}}{x^{g_B}}\right)^{N_B}\right]^{\frac{N_{A_2}}{N_{A_1}}} \mod p$.

7. Bob chooses a random message $M$ and sends $C = f(M, K'_A)$ to Alice.

8. Alice sends back $Y = f^{-1}(C, K_A)$ to Bob.

$c = f(m,k)$ is a function known to both Alice and Bob, where $m$ is the input, $k$ is the key and knowing $c$ does not reveal the key used. $f$ may be an encryption function using a secret key and $f^{-1}$ is the decryption function.

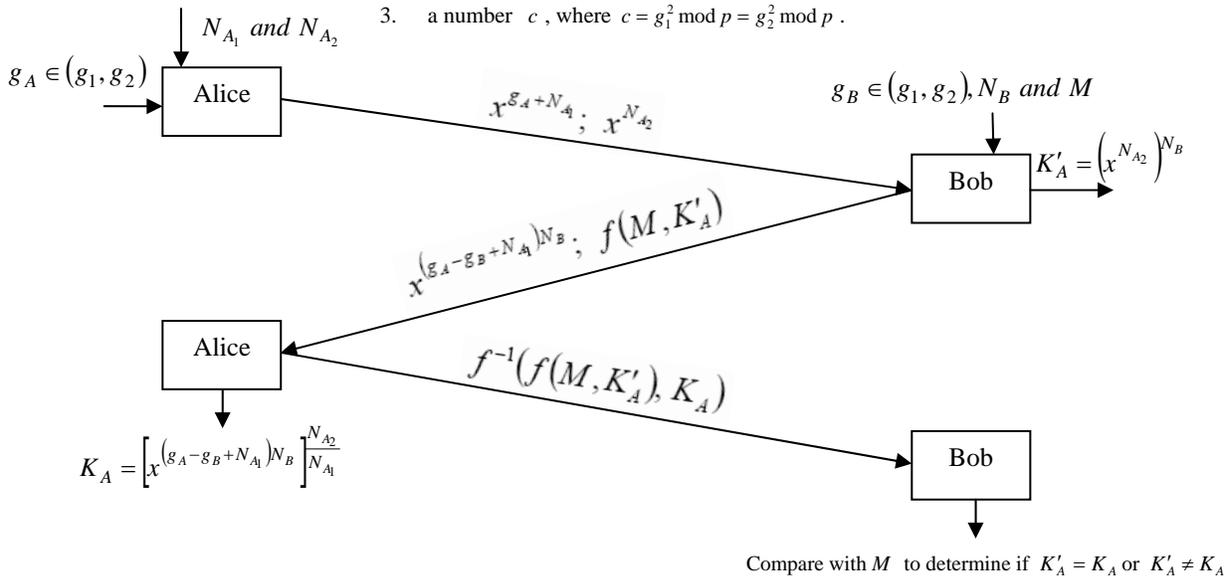

**Illustration of proposed algorithm to achieve oblivious exchange of encryption key (all computations performed in $Z_p$).**

Two cases arise from the above sequence, namely $g_A = g_B$ and $g_A \neq g_B$. If $g_A = g_B$ then $K'_A = K_A$, else $K'_A \neq K_A$. Hence, Bob receives $K_A$ with probability one-half. Steps 7 and 8 help Bob check if he has $K_A$ by comparing $Y$ and $M$.

Similarly, exchange of key $K_B$ takes place from Bob to Alice.

Define states,

$$U_b = \begin{cases} K, & \text{if Bob received } K_A. \\ \overline{K}, & \text{if Bob did not receive } K_A. \end{cases}$$

where, $K \in Z_p$ and $\overline{K}$ is the bitwise complement of $K$. $U_a$ is similarly defined.

In order to prevent cheating by either party, Alice sends $U_a \oplus S_A$ to Bob and Bob sends $U_b \oplus S_B$ to Alice. Since, neither party knows other's state of knowledge of the secret key, this step does not provide either party with any knowledge of other's secret.

Finally, Alice and Bob exchange their secrets encrypting them using $K_A$ and $K_B$, respectively.

If at the last step, after Alice sends her encrypted secret to Bob, Bob was to cheat and not send his secret to Alice, then the fact that Bob cheated implies that Bob received $K_A$ and $U_b = K$ and that Bob had previously sent $U_b \oplus S_B = K \oplus S_B$. Alice can retrieve $S_B$ by computing $K \oplus S_B \oplus K = S_B$.

The probability, after the protocol is complete, that neither party knows other's secret key is one-fourth.

***Example***: Alice and Bob wish to exchange secrets $S_A$ and $S_B$. They agree upon $p = 23$, $x = 5$ and $c = 9$. Therefore, $\sqrt{c} \bmod p = g_1 = 3$ and $\sqrt{c} \bmod p = g_2 = 20$. Two cases arise beginning from step 2 of the algorithm. Let us examine them:

Case I: $g_A = g_B$

2. Alice chooses: $g_A = g_1 = 3$ and two random numbers $N_{A_1} = 5$ and $N_{A_2} = 15$.

3. Bob chooses: $g_B = g_1 = 3$ and $N_B = 17$.

4. Alice sends to Bob: $x^{g_A + N_{A_1}} \bmod p = 5^{3+5} \bmod 23 = 16$ and
$$x^{N_{A_2}} \bmod p = 5^{15} \bmod 23 = 19.$$

5. Bob sends to Alice:

$$\left(\frac{x^{g_A+N_{A_1}}}{x^{g_B}}\right)^{N_B} \mod p = \left(\frac{5^{3+5}}{5^3}\right)^{17} \mod 23 = \left(\frac{16}{125}\right)^{17} \mod 23$$

$$= (16 \times 125^{-1})^{17} \mod 23$$
$$= (16 \times 7)^{17} \mod 23 = 7$$

and computes for himself: $K'_A = \left(x^{N_{A_2}}\right)^{N_B} \mod p = 19^{17} \mod 23 = 21$.

6. Alice computes: $K_A = \left[\left(\frac{x^{g_A+N_{A_1}}}{x^{g_B}}\right)^{N_B}\right]^{\frac{N_{A_2}}{N_{A_1}}} \mod p = (7)^{\frac{15}{5}} \mod 23 = 21$.

Bob may encrypt a random message with the key that he has generated and ask Alice to decrypt it using her key to determine if he has $K_A$. Since Alice and Bob have chosen $g_A = g_B = 3$, then $K'_A = K_A = 21$. ($g_A = g_B = 20$ gives similar results.)

Case II: $g_A \neq g_B$

2. Alice chooses: $g_A = g_1 = 3$ and two random numbers $N_{A_1} = 5$ and $N_{A_2} = 15$.

3. Bob chooses: $g_B = g_2 = 20$ and $N_B = 17$.

4. Alice sends to Bob: $x^{g_A+N_{A_1}} \mod p = 5^{3+5} \mod 23 = 16$ and
$$x^{N_{A_2}} \mod p = 5^{15} \mod 23 = 19.$$

5. Bob sends to Alice:

$$\left(\frac{x^{g_A+N_{A_1}}}{x^{g_B}}\right)^{N_B} \mod p = \left(\frac{5^{3+5}}{5^{20}}\right)^{17} \mod 23 = \left(\frac{16}{12}\right)^{17} \mod 23$$

$$= (16 \times 12^{-1})^{17} \mod 23$$
$$= (16 \times 2)^{17} \mod 23 = 9$$

and computes for himself: $K'_A = \left(x^{N_{A_2}}\right)^{N_B} \mod p = 19^{17} \mod 23 = 21$.

6. Alice computes: $K_A = \left[\left(\frac{x^{g_A+N_{A_1}}}{x^{g_B}}\right)^{N_B}\right]^{\frac{N_{A_2}}{N_{A_1}}} \mod p = (9)^{\frac{15}{5}} \mod 23 = 16$.

In this case, Alice and Bob have chosen $g_A \neq g_B$, hence $K'_A \neq K_A$. ($g_A = 20$ and $g_B = 3$ yields similar results.)

In none of the cases can Bob can predict before hand what choice Alice has made, so the protocol remains fair.

*Security issues*: The protocol breaks down if Bob is able to compute both $\left(x^{N_{A_2}}\right)^{N_B} \bmod p$ and $\left[x^{(g_A - g_B + N_{A_1})N_B}\right]^{\frac{N_{A_2}}{N_{A_1}}} \bmod p$. We see that Bob can deduce $x^{N_{A_1}}$ and $x^{N_{A_2}}$, using which he may compute $\dfrac{x^{N_{A_2}}}{x^{N_{A_1}}} = x^y \bmod p$. Given $y = N_{A_2} - N_{A_1}$, deducing $y$ is a DLP. If we assume that some how Bob is able to deduce $y$, then in order for him to compute the ratio $\dfrac{N_{A_2}}{N_{A_1}}$, he still needs to know either $N_{A_1}$ or $N_{A_2}$, which is again equivalent to a DLP. Based on the assumption that a Discrete Log Problem is difficult to solve, the protocol remains secure.

## 3  One-out-of-two oblivious transfer

One of the most powerful primitives that have led to the invention of numerous cryptographic schemes is the one-out-of-two oblivious transfer. It may conceptually be described as a black box where Alice puts in two secrets, $S_1$ and $S_2$, such that Bob can only retrieve one of them while getting no information about the other. Bob is concerned that Alice should not know which secret he retrieved.

A situation may be such that a spy wishes to sell one out of two secrets that he possesses, while the buyer does not wish the spy to know which information he wants. In such a situation the 1-out-of-2 oblivious transfer can be employed. It is assumed that the party possessing the two secrets is willing to disclose one and only one of these to the other.

The procedure of choosing prime $p$, generator number $x$ and $c = g_1^2 \bmod p = g_2^2 \bmod p$ remains identical to that described before. However, this time Alice uses secret keys $K_1$ and $K_2$ to encrypt secrets $S_1$ and $S_2$, respectively. She announces to Bob that she is associating key $K_1$ with $g_1$ and key $K_2$ with $g_2$. With these initial conditions the protocol follows:

1. Alice secretly chooses $N_{A_1}$ and sends to Bob: $x^{g_1 + N_{A_1}} \bmod p$.

2. Bob chooses $g_B = g_1$ (if he wants secret $S_1$) or $g_B = g_2$ (if he wants secret $S_2$) and secret numbers $N_B$ and $N_{B_1}$.

3. Bob sends to Alice: $\left(\dfrac{x^{g_1+N_{A_1}}}{x^{g_B}}\right)^{N_B N_{B_1}} \bmod p$ and $x^{N_B} \bmod p$.

4. Alice chooses a number $N_{A_2}$ and sends to Bob: $\left[\left(\dfrac{x^{g_1+N_{A_1}}}{x^{g_B}}\right)^{N_B N_{B_1}}\right]^{N_{A_2}} \bmod p$.

5. Bob computes: $K_B = \left[\left(\dfrac{x^{g_1+N_{A_1}}}{x^{g_B}}\right)^{N_B N_{B_1} N_{A_2}}\right]^{\frac{1}{N_{B_1}}} \bmod p = \left(\dfrac{x^{g_1+N_{A_1}}}{x^{g_B}}\right)^{N_B N_{A_2}} \bmod p$.

6. Alice computes: $K_1 = x^{N_B N_{A_1} N_{A_2}} \bmod p$ and $K_2 = \left(x^{N_B(g_1-g_2+N_{A_1})}\right)^{N_{A_2}} \bmod p$.

7. Alice encrypts secret $S_1$ using $K_1$ and secret $S_2$ using $K_2$ and sends them to Bob.

From the above sequence we see that if Bob chooses $g_B = g_1$, then $K_B = K_1$ and if Bob chooses $g_B = g_2$, then $K_B = K_2$. Hence, Bob will only be able to retrieve one of the two secrets depending upon his choice, while Alice will not be able to determine which secret Bob has retrieved.

*Security issues*: In order for Bob to cheat, he needs to compute both $K_1$ and $K_2$. His best option is to determine one of the keys honestly and using that, try to deduce the other key. For instance, if Bob honestly computes $K_1$, then he will have access to $x^{N_{A_1}}$ and $\dfrac{x^{N_{A_1} N_{A_2}}}{x^{N_{A_1}}}$. But this does not provide him with any information about $N_{A_1}$ and $N_{A_2}$ which he needs to compute $K_2$. Similarly, he cannot calculate $K_1$ from $K_2$. The problem is again equivalent to efficiently solving a DLP.

## 4  Coin-Flipping Protocols

A couple may decide on which restaurant to go to or whether they should take a vacation or buy a car for their next anniversary, by tossing a coin. In this case flipping a coin is a trivial matter since both parties are present at the same place physically. However, problems arise when the participants are geographically separated over large distances. How are they supposed to fairly flip a coin when both of them cannot see the outcome simultaneously? Many business transactions require such an arrangement or a simple game of gambling over the Web may need a fair coin-toss. Numerous solutions exist for this purpose that employ cryptographic techniques of bit commitment [9, 10].

It turns out that any oblivious transfer scheme may be suitably modified to flip a coin, so can be the protocol for mutual exchange of secrets that we have presented. For instance, if Bob receives the same key as Alice then Bob wins the toss else Alice wins. After Bob declares the key he has computed, Alice replies if he won or lost and reveals all the variables that she had chosen which Bob can use to verify Alice's claim. Bob may not disclose any of the variables of his choice.

## 5   Zero-Knowledge Proofs

Introduced in 1985, zero-knowledge proofs are typically used to force malicious parties to behave according to a predetermined protocol. In addition to their direct applicability to cryptography, they serve as a good benchmark for the study of various problems regarding cryptographic protocols [11, 12, 13]. Here we discuss a protocol for a prover $P$ to convince a verifier $V$ that he possesses certain information without disclosing the actual information. We may formally describe the problem as the following: $P$ declares a $y$, such that $y = x^e \bmod p$, where $p$ is a prime and $x \in Z_p$. $y$, $x$ and $p$ may be global information. However, only $P$ knows the exponent $e$. For everyone else, determining $e$ is a DLP. The problem is for $P$ to convince $V$ that he knows the value of $e$ without disclosing it. The protocol may proceed as follows:

1. $P$ chooses a random integer $n$ and sends $X = (x^e)^n \bmod p = x^{en} \bmod p$ to $V$.

2. $V$ chooses a random bit $b$. If $b = 0$, $M = 0$; else he chooses a random $m$ and sets $M = x^m \bmod p$ and sends $(b, M)$ to $P$.

3. If $b = 0$, $P$ sends $n$ to $V$; else $P$ sends $Y = M^e \bmod p = x^{me} \bmod p$ to $V$.

4. When $b = 0$, $V$ verifies $X$ is equal to $y^n \bmod p = x^{en} \bmod p$. So he believes that $P$ knows the value of $n$. If $b = 1$, $V$ verifies if $Y$ is equal to $y^m \bmod p = x^{em} \bmod p$. So he is convinced that $P$ knows the value of $e$.

This is a single round of the protocol. Upon multiple rounds of the protocol, the probability of an imposter cheating the verifier decreases exponentially.

We see that an imposter who does not know $e$ will succeed with a probability of one-half in each round. This is because if $V$ starts communicating with an imposter $P'$ from round one, then when $b = 0$, $P'$ successfully completes the protocol, but when $b = 1$, then $P'$ will have to guess the value of $e$. Hence, after $t$ iterations, the probability of the verifier being cheated decreases to $2^{-t}$. The protocol is zero-knowledge because $P$ never sends $e$, but only uses it as an exponent. This makes it equivalent to Discrete-Log-Problem.

A zero-knowledge proof can be used for identification if the verifier knows the value of $e$, which acts like a password. The prover has to convince the verifier that he knows the password, without actually giving it out. This is because the verifier may be an imposter trying to determine the password by cheating.

# 6  Conclusion

Our algorithm opens up the possibility of development of oblivious transfer schemes using key exchange protocols. Academically, it appears that such algorithms should have preceded Rabin's protocol. It shows that there exist numerous variations on the implementation of OT protocols. Also, most OT schemes can be extended to coin flipping with minor modifications, in which case, only one sided transfer may take place and success or failure depends on the opposing party being lucky enough to deduce the key.

Our protocol is different from Rabin's protocol in the sense that the latter aims at obliviously transmitting the decryption key from the transmitter to the receiver whereas we establish a shared key between the transmitter and receiver with probability one-half. Higher exponents may be employed to generate transfer probabilities other than one-half. It turns out that the Diffie-Hellman protocol is a powerful primitive and can be used as a basis for implementing many cryptographic protocols that have been implemented via the RSA type transformations. This possibility had been overlooked.

## Acknowledgement

I sincerely thank William Perkins and James Harold Thomas for discussions and useful comments.